# Title

Membraneless organelles formed by liquid-liquid phase separation increase bacterial fitness

# Authors

Xin Jin[1]†, Ji-Eun Lee[2]†, Charley Schaefer[2]†, Xinwei Luo[1]†, Adam J. M. Wollman[3], Alex L. Payne-Dwyer[2,5], Tian Tian[1], Xiaowei Zhang[1], Xiao Chen[1], Yingxing Li[4,1], Tom C. B. McLeish[2], Mark C. Leake[2,5]*, Fan Bai[1,6]*

# Affiliations

[1] Biomedical Pioneering Innovation Center (BIOPIC), School of Life Sciences, Peking University, Beijing, China.

[2] Department of Physics, University of York, York, United Kingdom.

[3] Newcastle University Biosciences Institute, Newcastle, United Kingdom.

[4] Department of Medical Research Center, Peking Union Medical College Hospital, Chinese Academy of Medical Science & Peking Union Medical College, Beijing, China.

[5] Department of Biology, University of York, York, United Kingdom.

[6] Beijing Advanced Innovation Center for Genomics (ICG), Peking University, Beijing, China.

*Correspondence to:

Fan Bai, fbai@pku.edu.cn; Mark C. Leake, mark.leake@york.ac.uk

† These authors contributed equally to this work.

# Abstract

Liquid-liquid phase separation is emerging as a crucial phenomenon in several fundamental cell processes. A range of eukaryotic systems exhibit liquid condensates. However, their function in bacteria, which in general lack membrane-bound compartments, remains less clear. Here, we used high-resolution optical microscopy to observe single bacterial aggresomes, nanostructured intracellular assemblies of proteins, to undercover their role in cell stress. We find that proteins inside aggresomes are mobile and undergo dynamic turnover, consistent with a liquid state. Our observations are in quantitative agreement with phase-separated liquid droplet formation driven by interacting proteins under thermal equilibrium that nucleate following diffusive collisions in the cytoplasm. We have discovered aggresomes in multiple species of bacteria, and show that these emergent, metastable liquid-structured protein assemblies increase bacterial fitness by enabling cells to tolerate environmental stresses.

**MAIN TEXT**

**Introduction**

Liquid-liquid phase separation (LLPS) drives the formation of membraneless compartments in eukaryotic cells, such as P-granules, nucleoli, heterochromatin and stress-granules (*1–5)*, enabling concentrations of associated biomolecules to increase biochemical reaction efficiency, and protection of mRNA or proteins to promote cell survival under stress. The characteristic feature of reversibility of LLPS, that is often highly sensitive to several microenvironmental factors, offer potential benefits to cells in terms of dynamic compartmentalization of their complex cytoplasmic components in response to changes in cellular physiology, in the absence of membranes (*1-2)*.

Compared to eukaryotes, the bacterial cytoplasm is more crowded and in general lacks membrane-bound organelles (*6*). Nonetheless, the localization of its cellular macromolecules is highly organized and exhibits non-uniform spatial distributions (*7*). For instance, chemotactic receptors recognize cell curvature to localize to cell poles (*8*), and the pole-to-pole oscillatory systems help to define the cell division plane (*9*). Recently, a few reports identified LLPS as a mechanism for organizing biomolecular condensates in bacteria (*10–12*). Al-Husini *et al*. showed that RNase E in *Caulobacter crescentus* assembles biomolecular condensate BR-body through LLPS, similar to the structure of eukaryotic P-bodies and stress granules (*10*). The RNA polymerase (RNAP) clusters in *Escherichia coli* (*E. coli*) were reported to be liquid-like droplets which are assembled through LLPS (*12*). ParB and ParS in the bacterial DNA segregation system associate to form liquid-like condensates that is driven by LLPS, whereas the ATPase activity of the ParA motor is required to counter the fusion of ParB condensates and maintain well-separated ParB condensates (*11*).

Protein aggregates in bacteria has been identified for many decades, but their functional role remains a debate (*13*, *14*). Protein aggregates in bacterial were initially considered as a collection of mis-folded proteins resulted from detrimental intrinsic or environmental stresses (*15–17*). These protein aggregates showed slow and limited post-stress disaggregation and remained in the mother cell for many generations (*18*).

However, recent studies showed that protein aggregates contribute to the asymmetric division of bacteria, and shape population heterogeneity which increases bacterial fitness (*14*, *19*, *20*).

Recently, we revealed the existence of bacterial aggresomes, subcellular collections of endogenous proteins that appear as dark foci in the cell body (*21*). Different from protein aggregates formed by mis-folded proteins as reported previously, aggresomes are dynamic, reversible structures that form in stressed conditions and disassemble when cells experience fresh growth media (*21*) (Fig. S1A). Here, we exploit experiment and multiscale modeling in tandem to determine the molecular biophysics of their spatial and temporal control, in order to deduce their crucial role in bacterial fitness. In the light of recent constructive criticism in the LLPS field (*22*), we have been particularly careful in testing the liquid nature of the aggresome.

## Results
### Aggresome formation in bacteria

We initially screened three proteins (HslU, Kbl and AcnB) as biomarkers, according to our previous analysis of their abundance in aggresomes (*21*), and labeled each with fluorescent protein reporters at chromosome (Tables S1-3). All three accumulated in aggresomes, though with different degrees of colocalization to dark foci in brightfield images (Fig. S1B). We overexpressed HokB protein, a toxin causing cellular ATP depletion (*23*), to trigger aggresome formation in *E. coli* (Fig. S1C). Monitoring the proportion of cells containing distinct fluorescent foci from time-lapse epifluorescence microscopy indicated different clustering dynamics with respect to ATP depletion (Fig. 1A). HslU, which had the highest colocalization to dark foci and was the most rapid responder with respect to ATP depletion, showed distinct fluorescent foci in 50% of cells after 60 mins HokB induction, followed by Kbl (requiring 150 mins), while the formation of AcnB foci was significantly slower. Dual-color fluorescence imaging confirmed the colocalization between HslU, Kbl and AcnB and their differing dynamics of incorporation into aggresomes (Fig. 1B, Fig. S1D).

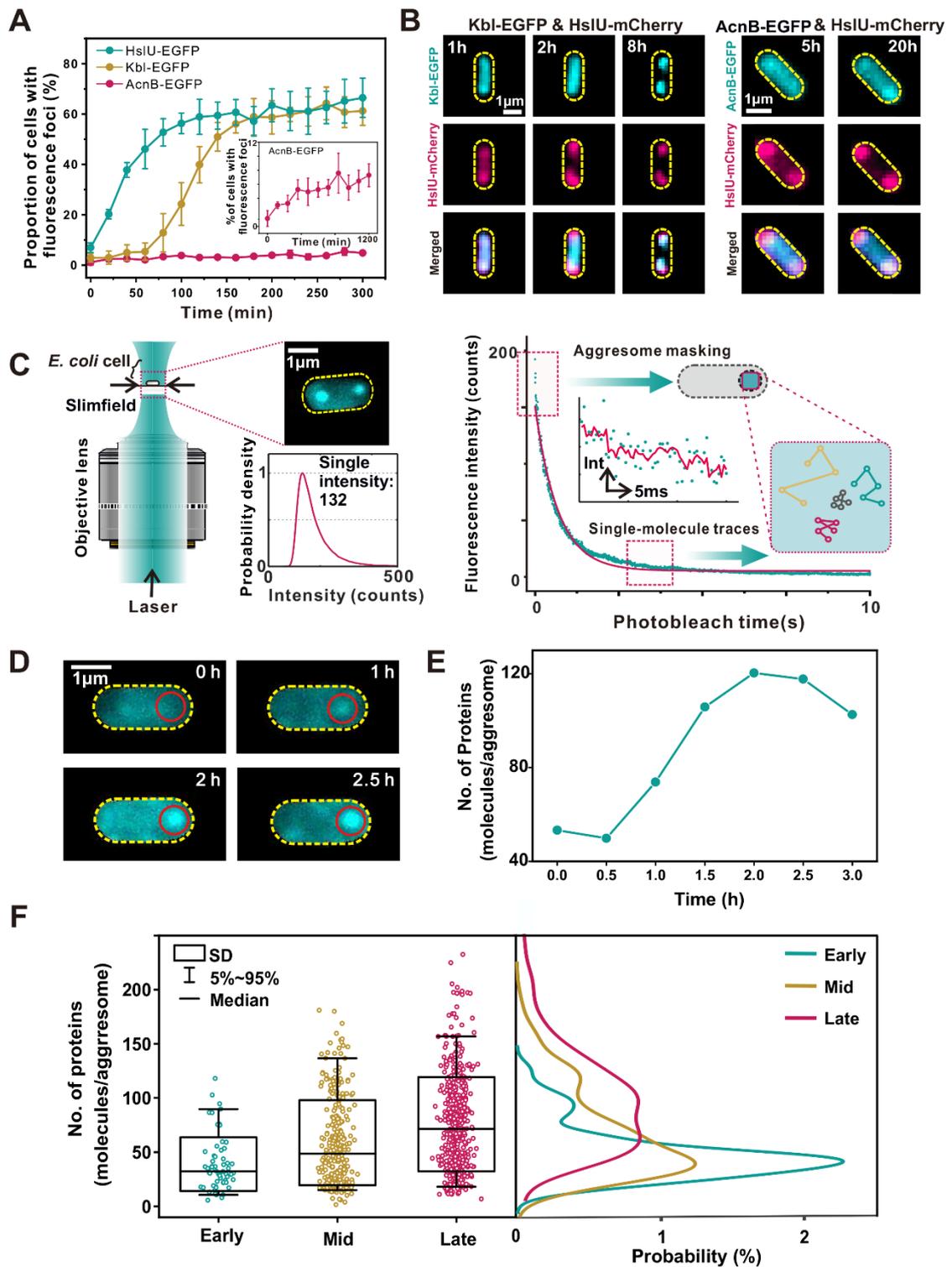

**Fig. 1. Aggresome formation in bacteria.** (A) Proportion of cells showing fluorescent foci as a function of HokB induction time. (B) Fluorescence images of dual-labeled strains indicating protein-dependent dynamics of cluster formation. (C) Left: Schematic of Slimfield microscopy used for rapid super-resolved single-molecule tracking. Inset shows the probability density function of detected single-molecule intensity values for EGFP *in vivo* (detected from 858 HslU-GFP labeled aggresomes taken from 370 different cells) with the vertical axis normalized semi-arbitrarily to 1 corresponding to the modal intensity value. This value corresponds to an intensity of 132 counts on our camera detector, which we denote as the characteristic integrated intensity for a single fluorescent EGFP molecule *in vivo*. Right: For detecting aggresomes, we used a frame average composed of the first five frames at the start of the photobleach of each image acquisition. For tracking single molecules, we looked into the end of photobleach processes. Inset: A representative step-wise photobleaching trace of a single EGFP molecule fitted by an edge-preserving Chung-Kennedy filter (*24*) (red line). (D) Time-lapse fluorescence images for strain HslU-EGFP during aggresome formation (red circle). The imaging interval is 30 mins. (E) Number of HslU-EGFP per aggresome from (D), SD error bars. (F) Left: box plots for number of HslU-EGFP per aggresome at different time points. Number of aggresomes N=62, 306, and 490 for early, mid, and late respectively. Right: Kernel density estimation (*25*) of the number of EGFP molecules per aggresome at the different HokB induction stages.

**Aggresome formation occurs through LLPS**

Since HslU showed the highest abundance and rate of response following ATP depletion we used it as the best biomarker from the three candidates to explore spatiotemporal features of aggresome formation with rapid super-resolved single-molecule tracking (Fig. 1C). We applied step-wise photobleaching analysis (*26*) to these data to determine the number of HslU-EGFP molecules visible inside and outside aggresomes in individual live cells (Fig. 1D). These data showed that the mean HslU-EGFP stoichiometry within aggresomes with respect to HokB induction time increases initially and then saturates, indicating demixing of HslU-EGFP from the surrounding cytoplasm in response to ATP depletion over a timescale of 1-3 hrs (Fig. 1E), that we categorized into 'early' (0.5 hr after induction, N=31), 'mid' (1 hr after induction, N=130), and 'late' (2-3 hrs after induction, N=209) stages (Fig. 1F).

To investigate the patterns of mobility for HslU in the cytoplasm, we photobleached the majority of HslU-EGFP and tracked the movement of fluorescent foci with only a few HslU molecules (Fig. S2A). We determined the apparent diffusion coefficient (*D*) of foci from the initial gradient of the mean square displacement (MSD) with respect to tracking time (Fig.2A, showing the spatial mapping of representative diffusive trajectories of HslU-EGFP proteins in the bacterial cytoplasm). To determine the molecular mobility of HslU inside aggresomes we tracked the fluorescent foci with only a single molecule, as confirmed by brightness and circularity (Fig. S2B-C). The mean diffusion coefficient for single molecule HslU-EGFP ($D_g$) across a population of cells decreased from $0.32 \pm 0.05 \mu m^2/s$ (± SE, early stage, number of tracks N=183), to $0.24 \pm 0.04$ $\mu m^2/s$ (mid stage, N=176), and then $0.19 \pm 0.02$ $\mu m^2/s$ (late stage, N=303) (Fig. 2B). Tracking single HslU-EGFP also revealed MSD maxima showing an aggresome diameter of up to a mean of 305 nm at late stage (Fig. S3A), which is consistent to within experimental error with single-particle tracking photoactivated localization microscopy (sptPALM) measurements (Fig. S3B-E).

We then corrected the apparent diffusion coefficient of these single-molecule tracks by subtracting the bulk diffusion measured for the whole aggresome itself ($D_a$), which was obtained by tracing the movement of the centroid of the aggresome, to yield the diffusion estimates for individual HslU molecules relative to its confining aggresome. The mean values of $D_a$ were $0.20 \pm 0.06$ µm²/s (early stage), $0.15 \pm 0.02$ µm²/s (mid stage), and $0.09 \pm 0.01$ µm²/s (late stage) (Fig. 2B). To check that these $D_a$ values are biologically reasonable, we modeled the aggresome diffusion ($D_a$) using the Stokes-Einstein relation for viscous drag associated with a sphere of diameter 305 nm that indicated a local cytoplasmic viscosity of 16.6 cP (1 cP = 1 mPa·s) (Table S4), a value broadly consistent with previous estimates on live *E. coli* (*27*, *28*). The mean values of $D_g$-$D_a$ were $0.12 \pm 0.07$ µm²/s (early stage), $0.09 \pm 0.05$ µm²/s (mid stage), and $0.10 \pm 0.03$ µm²/s (late stage), contrasted against $0.05 \pm 0.02$ µm²/s (greater than zero due to finite photon sampling) in separate experiments using identical imaging on single EGFP *in vitro* immobilized via antibodies to the glass coverslip surface (*29*) (Fig. S3F-G). The $D_a$ - $D_a$ of Kbl and AcnB (Table S5) is also higher than immobilized EGFP. Importantly, therefore, single HslU-EGFP inside aggresomes diffuse faster than immobilized EGFP at all stages. HslU molecules are thus definitively mobile inside aggresomes, consistent with a liquid state.

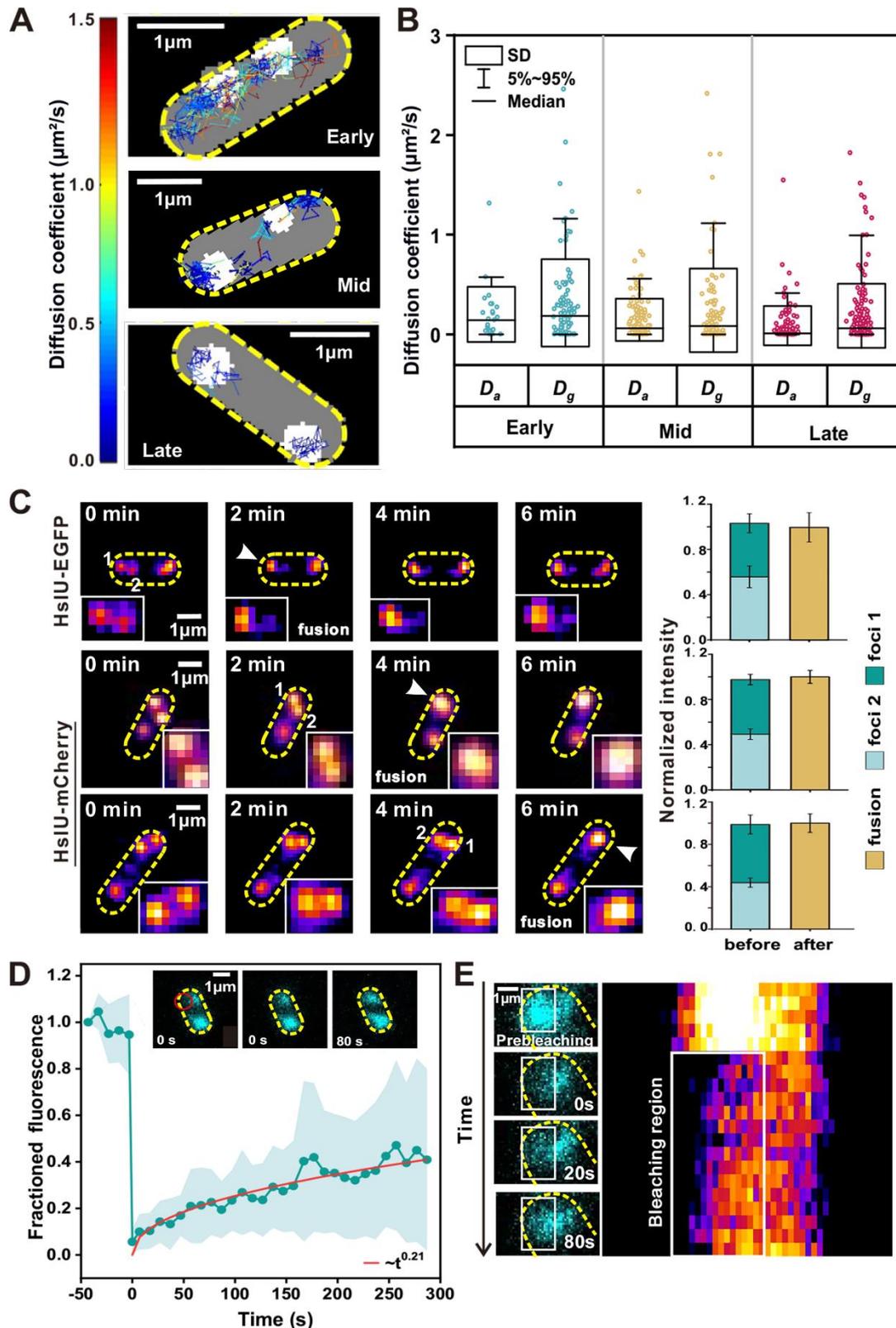

**Fig. 2. Aggresome formation occurs through LLPS.** (A) Representative tracking of HslU-EGFP foci during aggresome formation, color corresponding to diffusion coefficient (left bar). (B) Box plots of $D_g$ and $D_a$ of HslU-EGFP at different time points. (C) Left: Representative epifluorescence images of two aggresomes fusing. Right: Total fluorescence intensity of aggresomes before and after fusion, error bar represents the SD pixel noise. (D) Mean fluorescence recovery curve (cyan) defined

from half region of an aggresome indicating a mean recovery half time of 50 ± 9 secs and a heuristic power-law fit (red: ~$t^{0.21}$), with inset half-FRAP images of an aggresome showing fluorescence recovery, position of focused FRAP laser waist indicated (red). Error bounds show SD (number of aggresomes N=29, each from a different cell). (E) Zoom-in of aggresome showing fluorescence recovery with kymograph (right heatmap).

A common feature of LLPS is fusion of droplets to form larger, spherical compartments. To investigate this characteristic in aggresomes, we performed time-lapse epifluorescence microscopy at 2 min intervals to capture fusion events, using EGFP and mCherry reporters of HslU to ensure that this was not uniquely associated with just one type of fluorescent protein. As shown (Fig.2C and Fig. S4A) foci fusion often occurred at early- or mid-stage aggresome formation, with the fluorescence intensity of merged compartments tallying with the sum of intensities of individual foci to within measurement error. The fused aggresomes stayed in a circular spot for extended observation times of several minutes following fusion. 3D structured illumination microscopy (3D-SIM) and 3D stochastic optical reconstruction microscopy (3D-STORM) was used to determine the morphology of aggresomes at late stage (Fig. S4B-C). The mean aggresome diameter is 309 ± 51 nm (±SD, number of aggresomes = 100) for 3D-SIM measurement and 261 ± 32 nm (±SD, number of aggresomes = 37) for 3D-STORM measurement, with associated sphericity of 0.927 ± 0.04 and 0.86± 0.13 separately. Also, the circularity of the merged aggresome peaked at approximately 1 (Fig. S4D). These observations suggest that larger aggresomes may result from fusions of smaller HslU droplets, and support the hypothesis that LLPS drives bacterial aggresome formation. We also treated the cells at mid and late stages with 1, 6-hexanediol (HEX), which dissolves liquid-like condensates but not solid-like aggregates (*30*). As show in Figure S4E, 48% of aggresomes at mid stage disappeared (N=68) and 15% of aggresomes at late stage disappeared (N=100) after the addition of HEX.

To further test the liquid nature of aggresomes, we implemented FRAP measurements. By focusing a laser laterally offset approximately 0.5 µm from an aggresome center it was possible to photobleach EGFP content within approximately one spatial half of the aggresome, while leaving the other half intact, denoting this "half-FRAP". We then measured the aggresome fluorescence intensity at 10 sec intervals for up to several hundred seconds afterwards (Movie S1). Fluorescence in the bleached region recovered with a half time 50 ± 9 sec (±SD) to approximately

35% of the initial intensity at steady-state (Fig. 2D, kymograph shown in Fig. 2E). Similarly, we observed fluorescence loss in photobleaching (FLIP) for the initially unbleached half with increasing time after the focused laser bleach (Fig. S5A). Taken together, these results reveal that HslU-EGFP undergoes dynamic turnover within the aggresome over a timescale of tens of seconds.

To investigate turnover of HslU-EGFP between aggresomes and the surrounding cytoplasm we performed "whole-FRAP" for which a focused laser was centered directly on an aggresome, enabling us to photobleach all its EGFP contents while leaving the bulk of the remaining fluorescence in the cell, including that for other aggresomes, intact. We saw that the fluorescence of bleached aggresomes recovered with a half-time of $32 \pm 18$ sec, almost twice as rapidly as half-FRAP, though the recovery fraction of whole-FRAP (~10%) is much lower than that of half-FRAP (Fig. S5B-C). We also noticed that the rate of loss of fluorescence for the equivalent whole-FLIP was greater if we measured the loss from the whole cell area outside a bleached aggresome as opposed to just the fluorescence from a second aggresome localized to the opposite pole of the same cell, suggesting that short timescale turnover is limited to the local vicinity outside an aggresome. The whole-FRAP measurement indicates that there is a turnover of HslU molecules inside and outside aggresomes.

**Simulating aggresomes using an Individual-Protein-Based Model**

We developed an Individual-Protein-Based Model (IPBM) (*31*, *32*) to interpret the experimental observations by simulating collective dynamics of aggresome formation, protein turnover and dynamics as LLPS under thermal equilibrium (Fig. 3A). From our previous work, we know aggresome formation involves many types of proteins (*21*). Here we model this ensemble effect using an effective field approximation, denoting non-biomarker proteins that participate in aggresome formation as a general protein A while a specific aggresome biomarker protein is denoted B. In the first instance we used the experimental results from HslU-EGFP as protein B to establish a value for the interaction energy between A and B. We are also able to interpret the results from aggresome proteins Kbl and AcnB using the same model but with different interaction energies (Fig S7H).

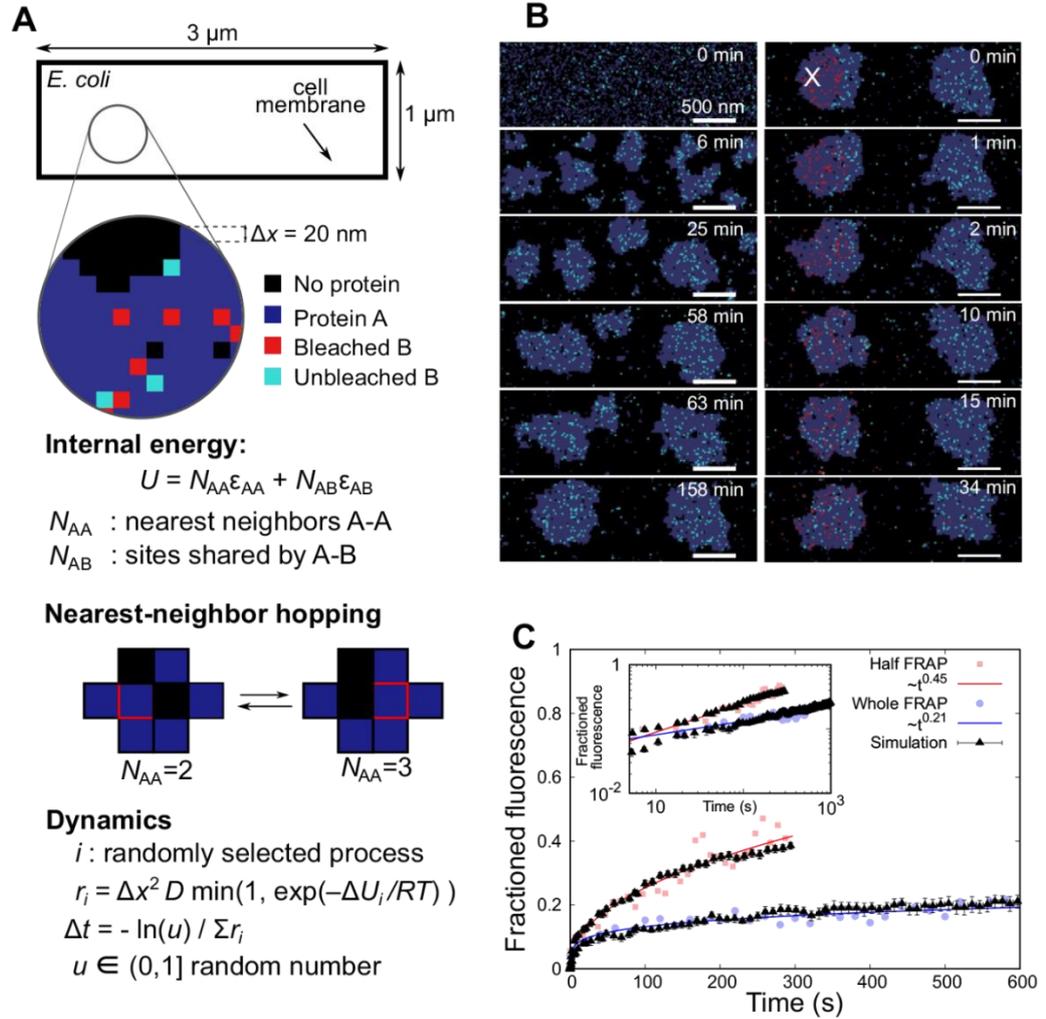

**Fig. 3. Simulating aggresomes using an Individual-Protein-Based Model.** (A) The cell is modeled as a 3x1 µm 2D rectangular lattice with regular 20 nm lattice spacing. Each lattice site can be either vacant, occupied by a single protein A (gray) or B (cyan), or by two proteins that are interacting (A and B). Protein A represents a general component in the aggresome that drives LLPS through an attractive nearest-neighbor interaction energy (optimized here to $\varepsilon_{AA}=2.2\ k_BT$) with protein B a fluorescently-labeled aggresome biomarker (such as HslU). The total internal energy is $U = N_{AA}\varepsilon_{AA} + N_{AB}\varepsilon_{AB}$, with $N_{AA}$ the number of A-A nearest-neighbor pairs, and $N_{AB}$ the number of sites shared by proteins A and B. Dynamics are captured using a stochastic Monte Carlo algorithm in which proteins move to nearest-neighbor sites with rate $r$ dependent on diffusivity $D$. (B) Simulations predict early-stage LLPS and coarsening by both Ostwald ripening and droplet fusion, resulting in two metastable droplets. Droplets form near cell poles and eventually merge but are stable within the experimental time scales. Proteins in a circular region of radius 350 nm are photobleached (red) after a time of approximately 140 minutes to simulate FRAP. (C) By quantifying the number of protein B molecules entering the photobleaching region we obtain simulated FRAP curves that obey scaling relationships that are in good quantitative agreement with whole- and half-FRAP experiments, here shown for HslU-EGFP ($R^2$ goodness-of-fit values of 0.756 and 0.917 respectively). 2,350 protein A molecules used equivalent to a volume fraction of 31.3%, with 400 protein B molecules.

The interactions and dynamics were modeled on a 2D rectangular lattice of 1×3 µm$^2$ (representing a geometrically simplified single *E. coli* cell) with grid spacing $\Delta x$=20 nm. Each lattice site may be vacant or occupied by a single protein A or B, or both. The internal energy is:

$$U = N_{AA}\varepsilon_{AA} + N_{AB}\varepsilon_{AB} \quad (1)$$

where $N_{AA}$ is the number of nearest-neighbor pairs of A, $\varepsilon_{AA}$ the corresponding pair-interaction energy, while $N_{AB}$ is the number of sites shared by the A and B proteins and $\varepsilon_{AB}$ is the interaction energy between A and B. We found that $\varepsilon_{AA}$=2.2 $k_BT$, which is above the critical value of ~1.8 $k_BT$ for LLPS (*33*), resulted in the best agreement to the experimental data. Within the framework of this model, it is expected that LLPS triggered by ATP depletion may be modeled by a change of the interaction parameter from $\varepsilon_{AA}$<1.8$k_BT$ to $\varepsilon_{AA}$>1.8$k_BT$.

We modeled the molecular dynamics using a (stochastic) Monte Carlo process where proteins hop to nearest-neighbor sites with an attempt frequency (*34*)

$$v = v_0 min(exp(-\Delta U/kT), 1) \quad (2)$$

where $v_0$=$D/\Delta x^2$ depends on the bare diffusivity $D$ in the absence of interactions, and $\Delta U$ is the energy difference due to the hop of a protein into a new local environment. At the interface a hop of B into or out of the aggresome is associated with a mean diffusivity $(D_{B,in} + D_{B,out})/2$. The time step for each hop depends on the sum, $S$, of all possible rates in Eq. (2) as $\Delta t = -\ln(u)/S$, with $0<u\leq 1$ a random number.

We generated representative LLPS behaviors (Fig. 3B and Movie S2), showing early-stage small clusters that coarsen according to a 1/3 power scaling with time as expected from Ostwald ripening and Brownian coalescence (Fig. S6A-D) (*35*). The key observation addressed by the time-dependent behavior is the qualitative feature of two droplets of comparable stoichiometry to experimental data forming at cell poles (Fig. 3B and Movie S2) over a timescale of 1-2 hours relevant to HokB induction times that only merge at much longer time scales. These model-interpreted observations suggest: (i) the long time scale does not necessarily point at the need to overcome a high free-energy to nucleate aggresomes (*36*), and (ii) that the observation of two aggresomes is a consequence of the elongated topology of the cell without a need of any activated processes; if we have the nucleoid in the center of the cell in our simulations that may further stabilize the two-droplet morphology (Movie S3).

The same model without change of parameters was able to reproduce simultaneously both the whole- and half-FRAP experiments (Fig. 3B-C). Here, we simulated LLPS using IPBM for approximately 140 mins prior to photobleaching a sub-population of protein B molecules localized to a laser bleach zone (approximated as a circle of radius 350 nm), after which the IPBM simulation continued and we monitored recovery as the number of unbleached protein B molecules in the laser bleach zone. In half-FRAP, fast recovery (~1 min) is governed by diffusion of unbleached B inside aggresomes, absent in whole-FRAP. At ~2 mins, recovery is more dependent on diffusion of unbleached B from the cytoplasm immediately surrounding aggresomes. At longer times, B from the opposite pole can reach the bleach zone and influence fluorescence recovery. We found that the experimental FRAP data for other aggresome proteins of Kbl and AcnB tried in our initial screen showed half-FRAP recovery to be consistent with a 1/2 power law (Fig. S7E). Therefore, we collapse all experimental and simulation data onto half-FRAP master curves (Fig. S7F-H). The differences in the offset of the whole-FRAP curves for the three types of proteins (Fig. S7H) seems to originate from differences in the binding energies of the respective proteins to the aggresome (Fig. S7F) and/or variations in their mobility in the cytoplasm (Fig. S7G). Our straightforward theoretical framework demonstrates that relatively simple protein physics – energetic interaction of diffusing proteins under local thermal equilibrium characterized by a simple and coarse-grained pairwise interaction energy – accounts not only for the conditions of aggresome formation in the first place, but also for complex, emergent biological features in their spatial localization and their dynamics and kinetics of molecular turnover.

**Aggresome formation increases the bacteria fitness**

Finally, we investigated if aggresomes confer physiological advantages to bacteria. We noticed that aggresomes are not unique to *E. coli*; by prolonged stationary culturing, we observed them in all eight of the other Gram-negative species we investigated (Fig. S8A). We performed large-scale screening of small chemical compounds and the *E. coli* gene knock-out library, and found three representative conditions, comprising media supplementation with MOPS (*37*) and deletions *ΔsdhC* and *ΔnuoA* (*38*), that disabled aggresome formation after prolonged stationary culturing (Fig. 4A-B).

MOPS has been reported as an osmolyte and protein stabilizer (*37*); in a wild type strain it penetrates the cell and stabilizes proteins from aggregation even if cytoplasmic ATP is depleted (*37*, *39*). Both *sdhC* and *nuoA* encode genes for the respiratory chain, where *sdhC* encodes the membrane-anchoring subunit of succinate dehydrogenase, and *nuoA* encodes NADH-quinone oxidoreductase subunit A. In respiration-impaired mutants, *ΔsdhC* and *ΔnuoA* strains, aggresome formation was disabled by inhibiting intracellular ATP decrease (Fig. 4C).

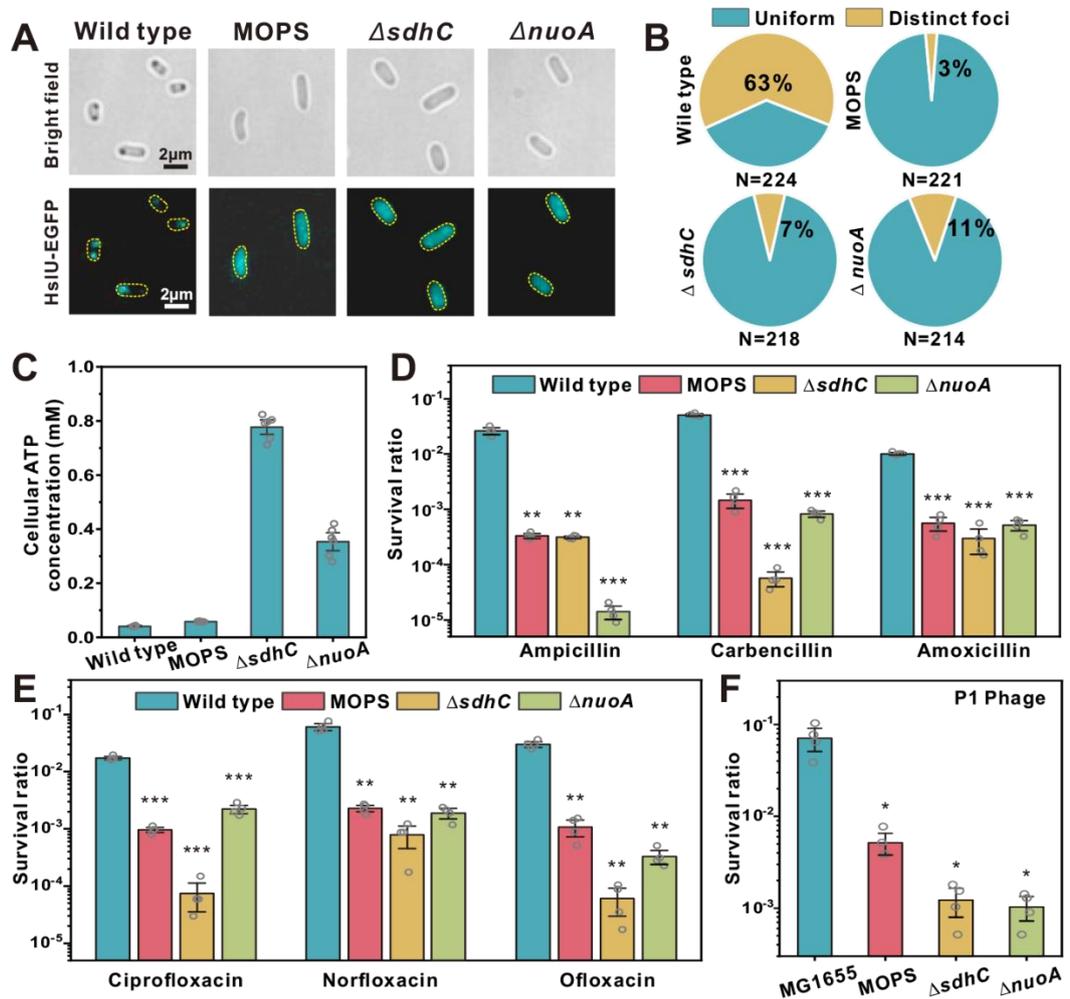

**Fig. 4. Aggresome formation promotes cellular survival under fierce stresses.** (A) Brightfield and fluorescence images of MG1655, MG1655 (MOPS), *ΔsdhC* and *ΔnuoA*, (B) proportion of cells showing HslU-EGFP foci, and (C) average cellular ATP concentration in different strains, after 24 hrs culture. Cell survival rate (log scale) after 4 hrs (D) β-lactam (E) fluoroquinolone, and (F) P1 phage infection, Multiplicity of Infection (MOI)=100. (Unpaired Student's t-test against wild type, error bar indicates SE, *p value < 0.05; **p value < 0.005; ***p value < 0.0005.)

We compared the survival difference between wild type, knockout strains *ΔsdhC* and *ΔnuoA*, and wild type supplemented with MOPS after stationary culturing. Interestingly, we observed that under various antibiotic treatments, the wild type strain with successful aggresome formation showed a higher persister ratio in comparison to strains in which aggresome formation was inhibited (Fig. 4D-E), indicating an increased fitness. We observed the same survival advantage in wild type when challenging bacteria with phage invasion (Fig. 4F). We also defined the survival ratio of these strains in early stationary phase, when no strains exhibited aggresomes, and found that all the strains have similar tolerance in early stationary phase (Fig. S8C-E). Taken together, we conclude that aggresome formation through LLPS promotes bacterial survival under a range of fierce stresses. This is consistent with our previous hypothesis (*21*): the formation of aggresomes sequester numerous proteins which are vital for cellular function, leading to shutdown of different biological processes and the cell enters a dormant state.

**Discussion**

How bacterial cells detect stresses and make their response specific and timely is an important question. A previous work has shown that the physical state of bacterial cytoplasm dramatically changes from fluid-like to glass-like as cells shift from metabolically active to dormant states (*40*). Here, we have established that aggresomes are membraneless liquid droplets hundreds of nm in diameter that phase separate following clustering of diffusing proteins under thermal equilibrium in the bacterial cytoplasm. These findings imply that bacterial cells potentially harness phase transition and separation processes in their stress-adaptive system and post-stress resuscitation. The phase transition and separation are reported highly sensitive to changes in certain parameters, such as pH, temperature, salt and molecular interactions.

Aggresomes contain numerous types of endogenous proteins. The three proteins selected in our study were incoporported into aggresomes over different characteristic timescales, indicating that the composition of aggresomes changes dynamically with time. Meanwhile, the diffusion coefficient of proteins (HslU) in aggresomes continued to decrease, which implies that aggresomes undergo some degree of gradual solidification. Thus, aggresomes induced by ATP depletion are highly heterogeneous in terms of composition and physical state. Kroschwald *et al*. found that the Pub1 stress granule condensates induced by different environmental stresses in yeast exhibit different material properties, which determine the rate of dissolution and the requirement of disaggregases (*41*). Giraldo *et al*. reported that RepA-WH1 amyloids in *E. coli* exhibited two different states, compact particles and a fluidized hydrogel, which show different toxicity (*42*). We speculate that bacterial cells tune the material composition and properties of aggresomes in response to different stress intensity and duration, and to set the strength of stress response and the length of recovery. This also provides a potential explanation to reconcile the controversy in the biological function of protein aggregation.

Protein aggregates in bacteria can be induced by diverse stresses, for instance, starvation, heat shock, oxidative stress and even antibiotic treatment (*21, 43-47*). A careful reassessment of the composition and physical state of bacterial protein aggregates or condensates is needed to explore their detailed biological function under stresses. In this work, we have demonstrated that LLPS serves as a highly reversible and sensitive way to compartmentalize cell cytoplasm in the absence of membranes. Our modeling indicates no requirement of external free energy for aggresome formation. A possible explanation could lie in recent work which demonstrated that ATP not only energizes processes inside cells but also acts as a hydrotrope to increase specific protein solubility (*48*); decreasing cellular ATP may thus act to favor LLPS. This method of regulation using ATP might be an important and recurring theme in several other biological processes that utilize LLPS.

## Materials and Methods

### Bacterial Strains, Phage and Plasmid Construction

All strains used in this study are indicated in Table S1, plasmids in Table S2, and primers in Table S3. Wild-type strain MG1655 was a gift from Yale Genetic Stock Center. P1 phage was a gift from Dr. Xilin Zhao, Xiamen University, China. Strains containing chromosomal *geneX-egfp/mcherry* translational fusion or single geneX knockout mutants were constructed by λ red mediated gene replacement (*49*). For fluorescent protein fusion strains: the targeted fluorescent protein (EGFP or mCherry) fragment was amplified and inserted to replace the stop cassette of the selection-counter-selection template plasmid (pSCS3V31C). Then the linker-FP-Toxin-CmR fragment was amplified from the template plasmid with homology arm complementary to the flanking sequences of the insertion site on the chromosome, before being transformed into electrocompetent cells with induced recombineering helper plasmid (pSIM6) (*50*). After 3-5 hrs recovery, transformed cells were plated on selection plates containing chloramphenicol (25 μg/mL). The Toxin-CmR cassette was then removed from the chromosome by another round of λ red mediated recombination using a counter selection template. Finally, cells were plated on counter selection plates containing rhamnose to activate the toxin (*51*). For knockout strains, we used the Keio cassette flanked by FRT sites in the Keio collection (*52*) to replace the certain gene on the chromosome.

The pBAD::*hokB* plasmid was transformed by electroporation into certain strains to generate *hokB* overexpressing strains. The *hokB* PCR product was amplified from

MG1655 and inserted into pBAD/myc-His A vector at *Nco I* and *Hind III* sites by the *in vitro* Gibson Assembly method. The ampicillin resistant gene was replaced with chloramphenicol resistance cassette. All strains were grown in Luria Broth (LB). We determined the cell doubling times of all strains generated to compare against their wild type equivalents (Table S1).

**Microscopy**

*Time-lapse microscopy*

Standard brightfield and epifluorescence imaging were performed on an inverted microscope (Zeiss Observer Z1, Germany). Illumination was provided by solid-state lasers (Coherent OBIS, USA), at excitation wavelengths 488 nm for EGFP and 561 nm for mCherry. The fluorescence emission signals of cells were imaged onto an EMCCD camera (Photometrics Evolve 512, USA). GFP ($\lambda_{em}$=500-550 nm) and mCherry ($\lambda_{em}$=604-640 nm) Zeiss filter sets were selected for each fluorophore according to their excitation and emission spectra.

Aggresome formation time-lapse imaging was performed using the Flow Cell System (Bioptechs FCS2, USA). Cells from the early stationary phase were washed with PBS using a centrifugation and were tipped on a gel-pad containing 2% low melting temperature agarose and 0.1% arabinose. Cells were then observed under brightfield and epifluorescence illumination at 37 ºC.

Home-written MATLAB software was used to detect fluorescent foci in time-lapse images. Cells were segmented according to brightfield images. Then, the fluorescence distribution inside the cells was analysed to determine whether there is a fluorescent foci. We used a 3×3 pixel (0.47×0.47 µm$^2$) kernel search for distinct fluorescent foci pixel-by-pixel inside each cell. If the foci met the two following conditions at the same time, we defined it as a fluorescent 'focus'. First, the mean fluorescence of the focus is higher than 1.5 times the mean fluorescence intensity of the whole cell. Second, the mean fluorescence of the foci is higher than 1.5 times of the mean local background intensity located in a 2 pixel ring area around a candidate focus. For brightfield black foci detection, cells were segmented first (excluding the white edges around each cell body image).Then we detected the black foci in the cells based on the gray value and area. If the gray value of a pixel is less than 95% of the average gray value of the whole cell, the pixel will be listed as a candidate of black foci. After the connected domain measurement, if the area is larger than 4 pixels, it will be defined as a black focus. The colocalization analysis was performed by analyzing the Pearson correlation coefficient of EGFP and mCherry fluorescent foci using Fiji plugin JACoP (*53*).

*Slimfield microscopy*

Microscopy was performed utilizing narrow epifluorescence excitation of 10 µm full width at half maximum (FWHM) in the sample plane to generate a Slimfield excitation field in the sample plane (*54*). EGFP was excited by a linearly polarized 488 nm wavelength 50 mW laser (Coherent OBIS, USA) attenuated to approximately 7 mW at the sample with average excitation intensity 4.6 kW/cm$^2$. To generate a circularly polarized beam, the laser was directed through a λ/4 waveplate and then to a dual-pass green/red dichroic mirror centered at long-pass wavelength 560 nm. Fluorescence emissions were captured by a 1.49 numerical aperture (NA) oil immersion objective lens (Nikon Apo TIRF, Japan) and passed through a bandpass emission filter with a 25 nm bandwidth centered at 525 nm (Chroma Technology, USA). Fluorescence emissions were imaged onto a Photometrics 95B CMOS camera, magnified to 50 nm/pixel. For the sample imaging protocol, a brightfield image was first acquired for 2 frames, and then we acquired EGFP images by the 488 nm excitation laser beam using a 5 ms exposure time per frame for 2,000 consecutive frames. For aggresome formation, the sample preparation used was the same as that in time-lapse microscopy.

*Single-particle tracking photoactivated localization microscopy (sptPALM)*

PALM images were acquired with a Zeiss Elyra TIRF imaging system bearing a 63× oil immersion objective (NA 1.46) (Zeiss Alpha Plan-Apochromat M27, Germany) and standard quad-band filters (Zeiss LBF 405/488/561/642 & BP 420-480 + BP570-640 + LP740) at a delivery angle of approximately 66°. Green and red channels were split using a long-pass dichroic (Zeiss SBS BP 490-560 + LP 640) onto the standard sCMOS sensors at 96.8 nm effective pixel size. Late-stage aggresomes were induced in a strain labelled with green-red photoswitchable HslU-mMaple3. Prior to switching, mMaple3 was imaged in the green channel under 488 nm wavelength excitation (4 mW ~ 0.1 kW/cm$^2$). Simultaneous photoswitching and excitation of converted mMaple3 in the red channel were then stimulated by 405 nm (0.3 mW ~ 8 W/cm2) and 561 nm wavelength (20 mW ~ 0.5 kW/cm$^2$) lasers respectively at 5.0 ms exposure per 7.6 ms frame interval for 5,000 frames. Photoswitched foci corresponding to single molecules of HslU-mMaple3 were identified and tracked in post-processing using the same custom MATLAB scripts described below for HslU-EGFP. The mMaple3 tracks were confined within circular regions as expected (Fig. S3C-D) and their terminal MSD plateaus (Fig. S3E) yielded an average aggresome diameter of 375 ± 86 nm (mean ± SD, n=583 tracks in N=23 aggresomes, late stage).

*3D Structure Illumination Microscopy (3D-SIM)*

3D-SIM images were acquired on a DeltaVision OMX SR imaging system (GE Healthcare, USA) equipped with a 100× oil-immersion objective (NA 1.49) and EMCCD, which achieves imaging of samples at approximately 120 nm lateral and 300 nm axial resolution. Laser lines at 488 and 561 nm were used for excitation. The microscope was routinely calibrated with 200 nm fluorescent microspheres. Serial z-stack sectioning was carried out at 125 nm intervals. Images were reconstructed with the softWoRx 5.0 software package. Reconstructed SIM data was further analyzed using software Imaris 9.6.0. The diameter of an aggresome is the equivalent spherical diameter (the diameter of a sphere of equivalent volume as the aggresome). The sphericity is the ratio of the surface area of a sphere (with the same volume as the aggresome) to the surface area of the aggresome. For reference, the sphericity of fluorescent microspheres is 0.973 ± 0.013.

Cells from the early stationary phase were washed and resuspended with PBS. Arabinose (0.1%) was added to induce aggresome formation. Cells were collected after 3 hrs induction (37 °C) and fixed in 2% PFA solution for 15 mins at room temperature and 30 mins at 4°C.

*3D stochastic optical reconstruction microscopy (3D-STORM)*

Images were acquired on Nikon N-SIM imaging system (Nikon, Japan) equipped with a 100× oil-immersion objective lens (NA 1.49), which achieves imaging of samples at approximately 20 nm lateral and 50 nm axial resolution. 10,000 frames at 200 ms intervals were acquired with excitation at 561 nm wavelength with 30% laser power and continuous activation at 405 nm wavelength with 1% laser power for HslU-mMaple3.

Cells expressing HslU-mMaple3 from the early stationary phase were washed and resuspended with PBS. Arabinose (0.1%) was added to induce aggresome formation. Cells were collected after 3 hrs induction (37 °C) and fixed in 2% PFA solution for 15 mins at room temperature and 30 mins at 4°C. We strictly avoid light exposure during sample preparation.

*Fluorescence Recovery after Photobleaching (FRAP) experiment*

FRAP was performed using a Zeiss LSM 710 confocal microscope with a 63× oil immersion objective (NA 1.4) (Zeiss Plan-Apochromat DIC M27, Germany). For half-FRAP experiments, a circular region of interest (ROI) of about 0.65 μm diameter in a half region of one aggresome (roughly 0.5 μm away from the aggresome's edge) was bleached using a laser focus with 488 nm wavelength generated by an Argon ion laser using a scanning time (i.e. bleaching time) equal to 1.62 frames per second (equivalent to 620 ms). The half region of one aggresome was bleached followed by the monitoring of the time course (every 10 seconds, 35 cycles, the first 5 cycles was the pre-bleaching sequence, and the remaining 30 cycles was the post-bleaching sequence) of fluorescence recovery. For whole-FRAP experiments, a circular ROI of about 1 μm diameter in one aggresome was bleached using the same laser with a scanning time (i.e. bleaching time) equal to 1.62 frames per second (equivalent to 620 ms). One selected aggresome was bleached followed by the monitoring of the time course (every 20 seconds 35 cycles), the first 5 cycles was the pre-bleaching sequence, and the remaining 30 cycles was the post-bleaching sequence) of fluorescence recovery. The emission detection wavelength ranged from 490 nm to 585 nm. The fluorescence recovery in the bleached region of interest was recorded by the commercial software (Zen Black) that controlled the microscope.

**Foci Detection, Tracking, Stoichiometry and Aggresome Masks**

Fluorescent foci were automatically detected using a custom program written in MATLAB (*55*) enabling estimation of the number of proteins per aggresome and the apparent diffusion coefficients for the whole aggresome and individual aggresome proteins. The detection and tracking software objectively identifies candidate fluorescent foci above a signal-to-noise ratio (defined as SN) fixed at 0.4 by a combination of pixel intensity thresholding and image transformation to yield initial approximations for the intensity centroid to the nearest pixel. The sub-pixel refinement to the centroid of each focus was then determined using iterative Gaussian masking. The intensity was defined as the summed pixel intensity inside a 5 pixel radius circular ROI corrected for the local background taken as the mean pixel intensity included in a 17×17 square ROI centered on the centroid but excluding the inner circular ROI. Each candidate focus was then fitted subsequently to an unconstrained 2D Gaussian fit to determine the separate σ widths in x and y relative to the camera detector. The circularity of that focus was then defined as $\sigma_x/\sigma_y$.

Characteristic intensity distributions of single EGFP molecules (Figure 1C) were rendered as Kernel density estimation (KDE) (*25*). Distributions were determined from the tracked foci intensity distributions from the end of the photobleach process confirmed by overtracking foci beyond their bleaching to generate individual steps of the characteristic intensity. For this, we collated all the foci after 1/3 bleaching. By doing that, only single fluorophore molecules are detected, that enabled estimation of the characteristic brightness of a single EGFP molecule in a living cell. This was qualitatively compared to *in vitro* estimates from surface immobilized purified EGFP (*29*) (Fig S3F), which agrees within 89 %.

Stoichiometry was determined by implementing a linear-fit to the first four intensity values of foci in each track, using the straight line going to just one frame ahead from the first frame of each track to define the initial intensity of each track and dividing this by the characteristic intensity of single EGFP.

Aggresome masks were defined from Slimfield image data by using the same spot detection algorithm as above but instead using a 5 frame average at the start of the photobleach of each image acquisition, resulting in comparable numbers of aggresomes detected per cell as those from detection of the distinct fluorescent foci from slower sampled epifluorescence microscopy. These aggresome foci were then fitted with a 2D radially symmetrical Gaussian function. A circular mask for each aggresome was then set with the center at the Gaussian centroid using a diameter of 1.5× the sigma width of the fitted Gaussian, resulting in a range of diameters across all aggresomes detected of 3-12 pixels.

**Mobility Analysis**

The two-dimensional apparent diffusion coefficient relative to the camera detector for each track was calculated from the gradient $G$ from a linear fit to first four data points 4 mean square displacement (MSD) values with respect to tracking time interval, i.e. equivalent to time interval values of 5, 10, 15, and 20 ms, with the fit intercept with the zero-time axis constrained to pass at the through $L^2$ where $L$ is the equivalent two dimensional localization precision estimated previously to be approximately 40 nm (*54*). $D$ was then determined as $G/4\Delta t$, where $\Delta t = 20$ ms.

$Dg$ denotes the values of apparent diffusion coefficient associated with tracks that were associated with just a single EGFP molecule. We confirmed that these were single-molecule tracks by looking at their stoichiometry which is 1 to within experimental error (Figure S2B). The $Dg$ was calculated by fitting the first 4 MSD values from the foci tracks, as above, found at the end of photobleach process. The mean values of $Dg$ were 0.32 ± 0.05 µm²/s, 0.24 ± 0.04 µm²/s and 0.19 ± 0.03 µm²/s for early-, mid-, and late-stage respectively.

We then corrected the apparent diffusion coefficient of these single EGFP molecule tracks inside aggresomes by subtracting the diffusion due to the associated aggresome ($Da$). $Da$ was calculated by fitting the first 4 MSD values from the foci tracks found within just the first five image frames of the start of laser illumination (i.e. the image frames used to generate the aggresome mask). The mean values of $Da$ were 0.20 ± 0.06 µm²/s, 0.15 ± 0.02 µm²/s and 0.09 ± 0.01 µm²/s for early-, mid-, and late-stage respectively.

**Estimation of aggresome diameter and cytoplasmic viscosity from Slimfield data**

We estimated the lower limit for the aggresome diameter under early-, mid- and late-stage induction with HokB by modeling using the following method. We collated certain tracks within each early-, mid- and late-stage dataset that contained one EGFP molecule to within experimental error on the basis of the measured foci brightness values. We then generated the mean average MSD versus time interval relation across of the single EGFP tracks found after 500 frames of the start of the laser illumination.

We then modeled the mean average MSD for each as that of a confining circular domain of radius *r* (*56*). In a confined diffusion in a circular domain, MSD is:

$$MSD = \delta d^2(t) \geq r^2 \left(1 - 8 \sum_{m=1}^{\infty} exp\left[-r_{1m}^2 \frac{t}{\tau}\right] \frac{1}{r_{1m}^2(r_{1m}^2 - 1)}\right)$$

The *d* parameter corresponds to the time-dependent distance diffused by particles. In the long-time limit t » τ, the MSD converges to $r^2$ for a circular domain, that is the projection of a cylindrical confining volume, or a thin cross-section through a spherical body. The equivalent limit for the 2D projection of a spherical body is MSD → $4r^2/5$, which becomes appropriate when the whole body lies within the imaged volume, i.e., r is not significantly greater than the microscope's depth of field. We then equated the maximum observed mean average MSD value in each dataset to an estimate for the lower limit of *r* and the lower limit of the aggresome diameter as 2*r*.

Estimates for the cytoplasmic viscosity were determined by using the Stokes-Einstein equation (*57*):

$$D = \frac{kT}{6\pi \eta r}$$

D is the diffusion coefficient from 0.29 μm²/s, 0.26 μm²/s and 0.19 μm²/s for early-, mid-, late-stage respectively; $k_B$ is Boltzmann's constant; *T* is the absolute temperature; $\eta$ is the dynamic viscosity; *r* is the radius of the spherical particle (see Table S4).

**Determining the Number of Proteins per Aggresome and Number of Proteins per Cell**

The number of EGFP molecules per aggresome was determined by calculating the total integrated intensity in the area of each aggresome mask. Within the aggresome mask, we first subtracted the mean background level of wild type *E coli* cells (MG1655) from each pixel, and then we defined mean value intensity among all the pixels. This mean value was divided by the characteristic single fluorophore intensity value. After that, this value was multiplied by the aggresome mask area. The probability distribution for the number of proteins per aggresome of HslU-EGFP was rendered using Gaussian kernel density estimation (KDE) (*25*) (Fig1F). To calculate the number of EGFP contained in the whole of a cell (i.e. the copy number), we used a previously developed CoPro algorithm that utilizes numerical integration and convolution of pixel intensities (*55*). In brief, we can calculate the fluorophore density from each pixel in the images. The cell pool was modeled as uniform fluorescence over rod-shaped cell of various cell length (depending on cells) and 800 nm width, which consisted in 2D of a rectangle capped by a half-circle at either end (corresponding to the 2D projection of the 3D *E. coli* cell shape of a cylinder capped by two hemispheres). The auto-fluorescence background level was determined by calculating the integrated mean intensity of parental strain (MG1655). The camera detector noise background defined by randomly selected area out of cell regions (mean size of camera background area ≈ a 25×25 square ROI) was subtracted from the mean pixel intensity value.

**FRAP Analysis**

The intensity traces of the half-FRAP or whole-FRAP bleached region targeted by the laser bleaching pulse were recorded by ZEN Black program as a function of time. For a photobleaching correction (i.e. due to photobleaching occurring during each sampling timepoint after the original focused laser bleach), we selected three cells in each video which were not been targeted and generated the equivalent intensity trace for them. To estimate the correction factor at each time point, we calculated the total integrated pixel intensities for each cell, and normalized this by the total integrated pixel intensity for each cell for the first time point in the series. We then calculated mean normalized intensity value for each timepoint across these three cells. The correction factor was taken as the reciprocal of this mean normalized intensity at each time point. To generate a corrected intensity for the FRAP intensity values after the focused laser bleach we then multiplied these by the correction factor for each corresponding timepoint. This process was carried out in each video and then the recovery curve was normalized by the initial intensity. For a final recovery curve, all the normalized recovery curves were averaged to determine the mean values. Fluorescence loss regions in half-FRAP experiments were selected as the half-unbleached aggresome region using ImageJ and then corrected for photobleaching using the same method as above. Fluorescence loss regions in whole-FRAP experiments were extracted by masking out the bleaching region of either one aggresome at the opposite pole of the cell, or from the whole cell area out the bleach region, as appropriate, and corrected for photobleaching as above. For half-FRAP experiments, aggresomes which showed fluorescence loss from the half-unbleached region were indicative of HslU-EGFP molecule mobility between unbleached and bleached regions.

**HEX treatment**

Cells expressing HslU-EGFP from the early stationary phase were washed and resuspended with PBS. Arabinose (0.1%) was added to induce aggresome formation. Cells were collected at different time points and washed with PBS to remove arabinose. Then, the cells are then loaded onto the gel-pad containing 2% low melting temperature agarose. The gel-pad was prepared in the center of the Flow Cell System (Bioptechs FCS2, USA) as a small gel island and surrounded by flowing PBS (volume of the liquid medium: volume of gel-pad = 1:20). Cells were then observed under brightfield and epifluorescence illumination at 37 ℃. Images were taken before HEX treatment. Then, 20% HEX solution was added to the chamber. Images were taken at 2 mins interval for 30 mins.

**Cellular ATP mesurement**

The cellular ATP concentration was measured using BacTiter-Glo Microbial Cell Viability Assay (Promega, G8231) following the manufacturer's instructions. Intracellular ATP concentration was determined through normalizing ATP levels by cell number and single cell volume.

**Antibiotic Sensitivity Assay**

Overnight bacterial cultures were diluted by 1:100 into LB medium (for MOPS treatment, 40 mM MOPS was added to the LB medium) and cultured at 37°C, 200 rpm. Bacterial cultures taken from the indicated time points were diluted by 1:20 into fresh LB with the following concentrations of antibiotics as appropriate for each sensitivity assay: 100 μg/ml for ampicillin, 10 μg/ml for carbencillin, 150 μg/ml for amoxicillin, 1 μg/ml for ciprofloxacin, 5 μg/ml for norfloxacin and 5 μg/ml for ofloxacin, respectively. Then the culture was returned to the 37ºC shaker for another 4 hrs. Samples were withdrawn and appropriately diluted in LB medium and spotted on an LB agar plate for overnight culture at 37 ℃. Colony counting was performed on the next day. There were four replicates.

**P1 Phage Sensitivity Assay**

Overnight bacterial cultures were diluted by 1:100 into LB medium (for MOPS treatment, 40 mM MOPS was added to the LB medium) and cultured at 37°C, 200 rpm for the indicated time. Then, bacterial cells and P1 phages were added to LB medium (supply with 5mM $CaCl_2$ and 5mM $MgCl_2$ to activate infection, MOI=100). After 10 mins incubation on ice for adsorption, the cells were transferred to 37 °C for 20 mins to complete infection (phage DNA injection). Then sodium citrate (final concentration 10 mM) was added to the culture to deplete $Ca^{2+}$ to terminate the infection process. Bacterial cells were washed with LB supplemented with 10mM sodium citrate following centrifugation at 3,000g for 2 mins. Samples were diluted as appropriate in PBS and spotted on LB agar plates (supplemented with 10 mM sodium citrate) for overnight culturing at 37 °C. Colony counting was performed on the following day. There were four replicates.

**Statistical Analysis**

Statistical tests were performed with the commercial software packages SPSS 18.0 as indicated in the figure legends.

**Acknowledgments**
    **Funding:**
    National Natural Science Foundation of China (No. 31722003, No.31770925) to FB
    Engineering and Physical Science Research Council (EPSRC) (EP/T002166/1, EP/N031431/1) to M.C.L.
    Biotechnology and Biological Sciences Research Council (BB/P000746/1, BB/R001235/1) to M.C.L. and J.-E.L.
    Royal Society (IEC\NSFC\191406) to M.C.L., the EPSRC (EP/N031431/1) to C. S. and T. C. B. M.

    **Author contributions:**

F.B., M.C.L., X.J., J.-E.L. conceived the study; X.J., J.-E.L., X.L., T.T., X.Z., A.P-D. performed the experiments; J.-E.L., X.J., C.S. led the data analysis with help from A.J.M.W. and A.P-D.; C.S. led the modeling and theoretical study; X.J., J.-E.L., C.S., F.B., M.C.L. wrote the manuscript with inputs from all authors. We thank Grant Calder at the York Bioscience Technology Facility for assistance with confocal and PALM microscopy. We thank National Center for Protein Sciences at Peking University in Beijing, China, for assistance with 3D-SIM imaging.

**Competing interests:** The authors declare no competing interests.

**Data and materials availability:** All data is available in the main text or the supplementary materials.